# Driving with temperature the synthesis of graphene on Ge(110)

L. Persichetti[1], M. De Seta[1,*], A. M. Scaparro[1], V. Miseikis[2], A. Notargiacomo[3], A. Ruocco[1], A. Sgarlata[4], M. Fanfoni[4], F. Fabbri[5], C. Coletti[2], and L. Di Gaspare[1]

[1]*Dipartimento di Scienze, Università Roma Tre, Viale G. Marconi, 446- 00146 Rome, Italy*

[2]*Center for Nanotechnology Innovation @NEST, IIT, Piazza San Silvestro 12, 56127 Pisa, Italy*

[3]*Institute for Photonics and Nanotechnology, Via Cineto Romano 42, 00156, CNR-Rome, Italy*

[4]*Dipartimento di Fisica, Università di Roma "Tor Vergata", Via Della Ricerca Scientifica, 1- 00133 Rome, Italy*

[5]*NEST, Istituto Nanoscienze – CNR, Scuola Normale Superiore, Piazza San Silvestro 12, I-56127 Pisa, Italy.*

We systematically investigate the chemical vapor deposition growth of graphene on Ge(110) as a function of the deposition temperature close to the Ge melting point. By merging spectroscopic and morphological information, we find that the quality of graphene films depends critically on the growth temperature improving significantly by increasing this temperature in the 910-930 °C range. We correlate the abrupt improvement of the graphene quality to the formation of a quasi-liquid Ge surface occurring in the same temperature range, which determines increased atom diffusivity and sublimation rate. Being observed for diverse Ge orientations, this process is of general relevance for graphene synthesis on Ge.

[*]corresponding author:

Fax: +39 0657333430 E mail: monica.deseta@uniroma3.it (Monica De Seta)

# 1. Introduction

The last decade has been marked by extraordinary advances in the field of two-dimensional (2D) materials made possible by the discovery of graphene [1]. This category of materials is characterized by an unprecedented wealth of attractive and excellent properties and functionalities that can be further tuned by confinement effects, as in graphene nanoribbons [2, 3]. Therefore, technological application of graphene materials [4-9] is expected to revolutionize fields which are currently dominated by group-IV elementary semiconductors, i.e. Si and Ge. To unlock this potential, a critical and necessary breakthrough is the integration of graphene films and nanostructures with the complementary metal-oxide-semiconductor (CMOS) manufacturing process of semiconductors [10]. To date, the main obstacle to the integration and CMOS compatibility [11] is the metallic contamination introduced in the graphene during the metal-catalyzed deposition process. Furthermore, transferred graphene films suffer from several practical issues hampering their CMOS integration such as undesirable processing complexity, difficulty in obtaining high transfer yield, contamination by polymer residues and solvents used for transfer [12, 13]. Alternative synthetic approaches, such as the unzipping of carbon nanotubes typically used for obtaining sub-10 nm graphene nanoribbons and flakes [14-16], are also affected by residual contamination by metal nanoparticles originating from the growth process of nanotubes and, in addition, face severe challenges in terms of positional control over the substrate and width heterogeneity [17]. Thanks to the catalytic activity of Ge on gaseous carbon precursors combined with Ge carbides instability, chemical vapor deposition (CVD) of graphene on Ge or Ge/Si substrates represents a viable growth route [18-24] for obtaining metal-free, CMOS-compatible graphene. After the seminal works by Wang *et al*. [18] and Lee *et al*. [19] on Ge(110), most of the attention has been focused on the Ge(001) surface, this being, in principle, the most technologically relevant one. On this face, the growth of high-quality graphene has already been demonstrated [22, 25-29], as well as the synthesis by CVD of ultra-thin ribbons having widths narrower than 5 nm [30-34]. Experiments have also shown, however, that the Ge(001) surface under the graphene flakes or ribbons is severely faceted along {1,0,L} orientations [20, 22, 25, 30, 35, 36], consistent with the existence of several surface-energy minima around the Ge(001) face which favor faceting at high temperature [37, 38] or under strain [39]. The nanofaceting development questioned the suitability of this interface for further technological processing, and has led to renewed interest in graphene/Ge(110) for which the underlying Ge surface not only remains flat but also promotes the formation of a graphene single crystal [19]. Attention has been focused on the electronic properties of graphene grown on Ge(110) [24, 40] and on the interfacial structure with the Ge surface [41-44], revealing at the interface both structural motifs related to the (1$x$1) Ge termination [42] and a characteristic (6$x$2) Ge reconstruction [41, 43, 44] which is not observed for bare Ge(110) without graphene. Morphologically, the growth of uniaxially aligned graphene islands [45] merging into a uniform graphene film [46] has been observed, whereas no nanoribbon formation has yet been reported so far, in contrast to the Ge(001) substrate. In this paper, we systematically investigate the CVD growth of graphene on Ge(110) as a function of the deposition temperature close to the Ge melting point. By combining atomic force (AFM) and scanning tunneling microscopies (STM) with Raman and X-ray photoelectron spectroscopies (XPS), we find that, in our growth conditions, the structural quality of graphene films depends critically on the growth temperature and improves significantly by increasing the deposition temperature in the 910-930 °C range. Within the same narrow temperature range, previous surface science investigations of Ge(110) revealed the formation of a quasi-liquid surface adlayer [47], an evidence in agreement with the occurrence of the incomplete surface melting of Ge [48]. This suggests an intimate connection between the surface-melting discontinuity in the Ge phase diagram and graphene's quality triggered by the increased mobility and sublimation rate on the quasi-liquid Ge, favoring the healing of defects in graphene [49]. In addition, we find that the annealing temperature has also a strong impact on the morphology of the Ge(110) surface where wide terraces and a low density of steps is obtained performing a pre-growth annealing to 930 °C. Such a substrate morphology is potentially suitable for the growth of graphene nanostructures such as nanoribbons. On this pre-annealed substrate, CVD deposition below 900°C produces, over a mostly uncovered Ge surface, sparse ribbon nanostructures, having widths of ~5 nm and heights compatible with the out-of-plane spacing of graphene/Ge(110), although the obtained density is at the moment too low to unambiguously confirm their graphene composition by spectroscopic characterizations.

## 2. Materials and methods

Graphene films were grown on Ge(110) substrates (*n*-type Sb-doped, $n=10^{16}$ cm$^{-3}$) using a commercial CVD reactor (Aixtron BM). Ge substrates were cleaned *ex-situ* by multiple rinsing and drying cycles alternating isopropyl alcohol and de-ionized water and then loaded into the growth chamber evacuated to a low-pressure regime ($10^{-4}$ mbar). A gas mixture of H$_2$/Ar (200/800 sccm) was introduced in chamber, reaching the desired background pressure of 100 mbar which is kept constant during the whole process. The thermal process starts with the heating of the substrate from room temperature (RT) to the desired temperature. We used two graphite heaters for homogeneous heating over the Ge substrate and adopted a multi-step temperature ramp with a rate progressively decreasing close to the target temperature. The chamber is initially heated up to 800 °C at a fast rate of 4 °C/s. An intermediate rate of 1 °C/s brings the temperature up to 860 °C. Finally, a slow rate of 0.125 °C/s was used to achieve the desired annealing temperature $T_C$ which is kept stable for 5 min. At this stage, the growth is either started with the introduction of 2 sccm of CH$_4$ (in case the deposition temperature $T_D$ coincides with $T_C$) or the temperature is ramped down to $T_D$ with the same slow rate. The described procedure ensured accurate control over the temperature close to the Ge melting. Moreover, we verified on a sacrificial sample that temperatures 5° C higher than 930 °C resulted in a partial melting of the Ge substrate. After growth, samples were cooled down to room temperature in H$_2$ and Ar atmosphere.

The graphene samples were characterized by using X-ray photoelectron and Raman spectroscopies. The XPS measurements were carried out using a monochromatic Al $K_α$ source ($hν$=1486.6 eV) and a concentric hemispherical analyzer operating in retarding mode (Physical Electronics Instruments PHI), with an energy resolution of 0.4 eV. The carbon amount $\rho$ deposited on the samples was estimated from the $C_{1s}$ core level area intensity normalized to that acquired in the same experimental conditions on a commercial graphene monolayer (CGM) positioned next to the analyzed sample, i.e. $\rho = I_{C_{1s}}^{sample} / I_{C_{1s}}^{CGM}$ [27].

Raman spectroscopy was performed with a Renishaw *inVia* confocal Raman microscope using an excitation wavelength of 532 nm, a 100*x* objective and a laser spot size of 1 µm. For a quantitative analysis, the intensity ratios of the Raman bands were obtained from the integrated intensity of the fitting peaks. The sample morphology was investigated by scanning electron microscopy (SEM) (FEI Helios 600 NanolabDualBeam), AFM (Bruker Dimension Icon microscope) operating in Tapping Mode, and ultra-high-vacuum (UHV), room-temperature scanning tunnelling microscopy (VT Omicron) working in constant current mode. Finally, low-energy electron diffraction (LEED) was employed for assessing the crystalline quality of graphene. Conventional back-display (*c*-LEED) measurements were paralleled by high-resolution LEED using a monochromatic electron gun and a hemispherical electron analyzer. Reciprocal space is spanned by the vectorial quantity ***q*** which is the momentum transfer of the diffraction experiment. In the specular (elastic) conditions, the component of ***q*** parallel to the surface vanishes, thus defining the zero-*th* order diffraction peak. By scanning the polar angle $\theta$ (while keeping fixed the incidence and collection angles), the scattering vector component parallel to the surface is changed and a surface diffraction peak is observed for ***q***$_{//}$ matching a surface reciprocal lattice vector. The crystallographic direction probed on the surface is instead defined by the azimuthal orientation $\phi$.

## 3. Results and discussion

In a first set of experiments, three graphene samples were deposited at $T_D$ equal to 910, 920 and 930 °C, respectively, using for each sample an annealing temperature $T_C$ equal to $T_D$ (i.e. $T_C = T_D$) and a deposition time of 60 min. The small variations in the deposition temperature produced striking differences in the grown graphene in terms of quality, as evident from Raman spectra in Fig. 1, and morphology at different length scales probed by SEM, AFM and STM (Fig. 2). Note, however, that the total carbon amount $\rho$ measured by XPS data depends very slightly on the temperature (varying less than 15%), being close to 1 for all the samples (see Table 1). This information, together with the absence of extended bilayer regions in SEM images [Fig. 2(a-c)], indicates that a graphene monolayer covers the Ge substrate almost uniformly in all the samples.

| $T_D$ (°C) | $\rho$ (graphene monolayers) | $\Gamma_{2D}$ (cm$^{-1}$) | $I_{2D}/I_G$ | $I_D/I_G$ |
|---|---|---|---|---|
| 910 | 0.88 | 48 | 1.7 | 2.0 |
| 920 | 0.91 | 47 | 2.4 | 1.5 |
| 930 | 0.99 | 46 | 3.0 | 0.3 |

*Table 1. Quantitative analysis of XPS and Raman spectra.*

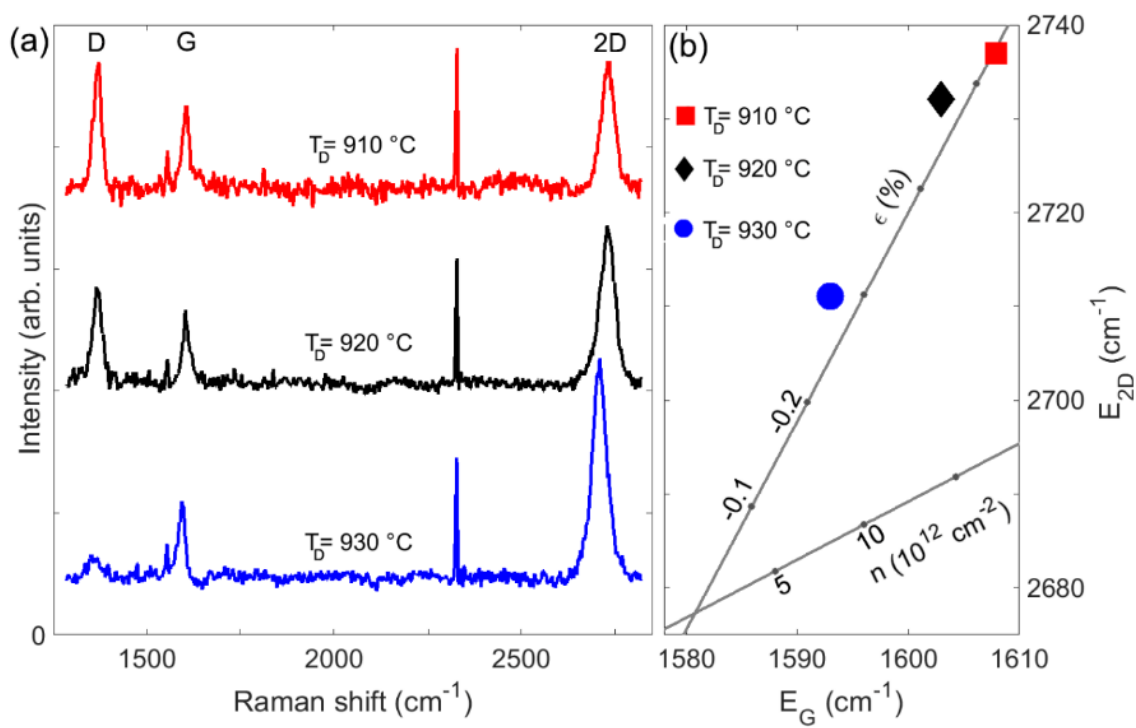

*Fig. 1. (a) Raman spectra of graphene samples grown on Ge(110) at different $T_D$ (=$T_C$). (b) Plot of the 2D- vs G-band energies. ε is the strain and n the charge density. Straight lines indicate $E_{2D}$ vs $E_G$ relationship for strained undoped (n= 0) and unstrained (ε= 0) n-doped graphene. The two lines cross at the expected 2D and G positions for suspended freestanding single-layer graphene (neutrality point).*

In Fig. 1(a), we observe the main Raman features of graphene, i.e. 2D and G bands respectively at ~2700 and ~1600 cm$^{-1}$ in all the spectra. In addition, the D peak originating from intervalley resonant scattering due to defects is also visible. At higher $T_D$ we observe a boost of the $I_{2D}/I_G$ intensity ratio produced by the increase of the 2D intensity (See Table I) while the width of the 2D peak is almost unaffected by $T_D$ variation. These observations indicate that, in the temperature range explored, an increase of $T_D$ as low as 20 °C produces a major improvement in the crystalline quality of the graphene film [50-52]. Accordingly, we observe that the $I_D/I_G$ ratio drops upon increasing $T_D$, suggesting an effective reduction in the defect concentration at high temperature.

By analyzing the diagram of the 2D *vs* G band energies [Fig. 1(b)], one can compare the strain and doping levels of the three graphene samples [53]. We find that the doping density is almost independent of the growth temperature $T_D$, this being, in line with previous observations [20], negligible on Ge(110). The strain, instead, shows a clear dependence on $T_D$. As the deposition temperature gets higher, it decreases from the value of -0.52 % at $T_D$= 910 °C to less than -0.3 % at 930 °C. The origin of this compressive strain is likely related to different thermal expansion coefficients (CTEs) of graphene [54] and Ge [55]. While at the deposition temperature the CTE of graphene is larger than that of Ge, below ~500 °C graphene CTE becomes rapidly smaller than that of the substrate: this means that a compressive thermal strain is built up in graphene during cooling, as observed by Raman spectroscopy [25]. Previous work showed that such a strain can be effectively reduced by formation of wrinkles in graphene [56]. By comparing the SEM images in Figs. 2(a-c), it is evident that, while no wrinkles are observed at $T_D$= 910 °C (i.e. where the strain is larger), a wrinkle pattern does appear at $T_D$= 920 °C and 930 °C. In the two samples the wrinkle edges have almost the same directions and form 120°angles. The tessellation areas defined by the wrinkle pattern widen markedly between 920 and 930 °C, showing at the highest temperature longer edges and a more homogeneous distribution over the surface. From AFM [Figs. 2(d-f)], it is also found that the wrinkle height increases with temperature, from ~1.5 nm at 920 °C to ~3 nm at 930 °C. Based on these observations, it is reasonable to hypothesize that the higher and more uniformly distributed wrinkles at 930 °C are capable of better relieving the thermal strain, so explaining the higher degree of relaxation measured by Raman spectra at this temperature.

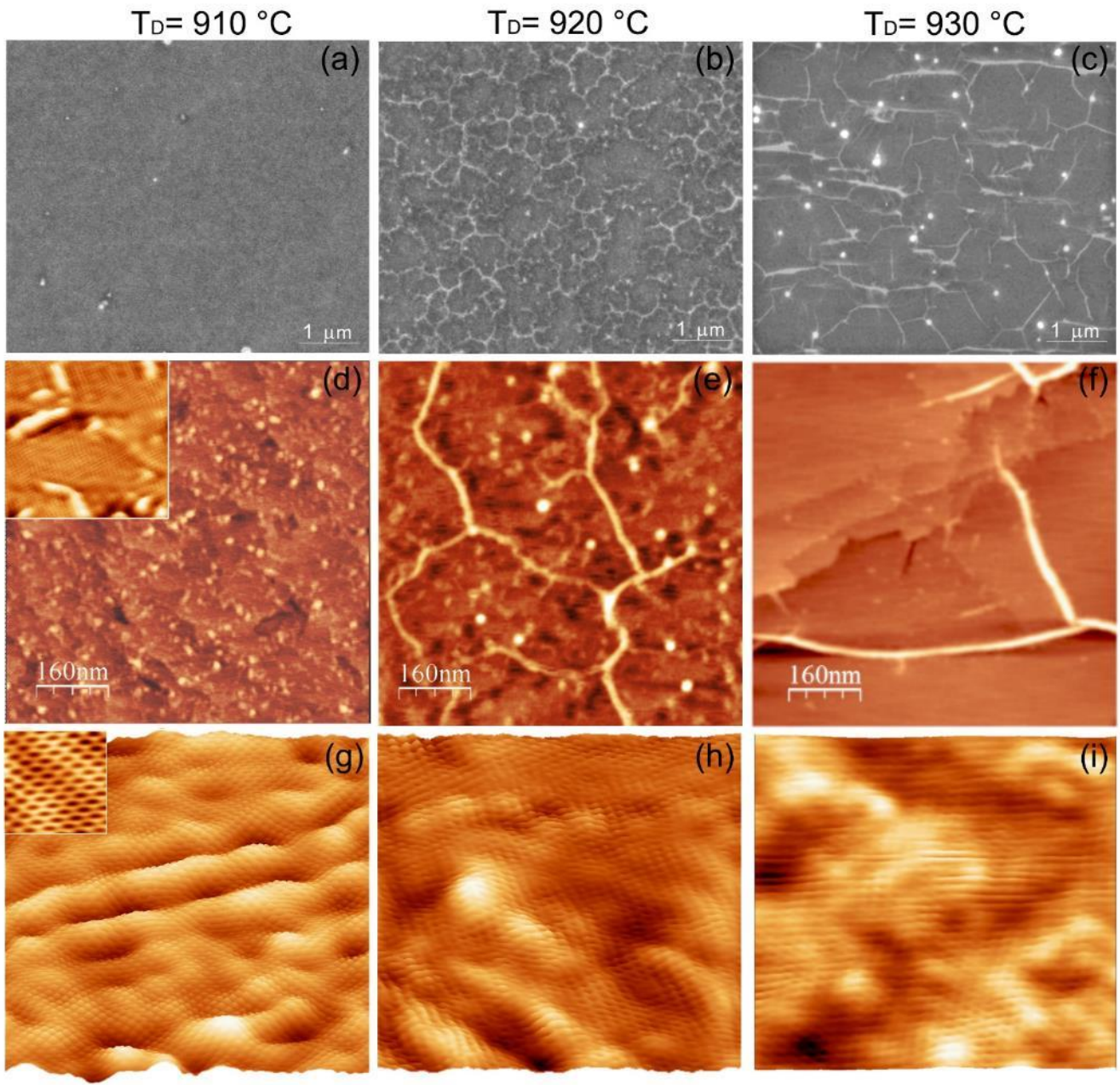

*Fig. 2. Morphological characterization of graphene films grown on Ge(110) at different $T_D$ (=$T_C$). (a-c) SEM images. (d-f) AFM images. The maximum of the z axis increases from (d) to (f) being, respectively, 0.9 nm, 1.8 nm and 3.2 nm. In the inset of panel (d), a STM image (50x 50) $nm^2$ is shown. (g-i) 3D STM topographies (8.2x 8.2) $nm^2$ displayed with the same z-axis range 0-0.12 nm. In the inset of panel (g), a (1.3x 1.3) $nm^2$ blow-up of the honeycomb lattice is displayed. STM images were taken with the following tunneling parameters U= 50 mV, I=1.0 nA.*

In addition to the presence/absence of wrinkles, the AFM topographies shown in Figs. [2(d-f)] reveal that the morphology of the Ge substrate also changes dramatically with temperature. Despite the fact that the typical terrace/step structure of the Ge surface is visible in all the samples, the terrace widths increase markedly with temperature, being <70 nm at $T_D$= 910 °C and reaching several hundreds of nanometers at 930 °C. In addition, in the samples grown below 930 °C we note the presence of ridges in AFM measurements. The density of these features is particularly high on the sample deposited at $T_D$= 910 °C [Fig. 2(d)]. Their structure is better clarified by STM imaging [inset Fig. 2(d)] which shows worm-like ridges and rounded bumps with a lateral size of a few nanometers and heights ranging between 0.3-0.5 nm. Moreover, it is evident that the graphene layer covers continuously the substrate, following the corrugation of the ridges. Among the ridges, graphene is not completely flat but shows a rippling pattern with a typical wavelength of ~1 nm. This ripple morphology present at low $T_D$ is better evident at higher magnification [Figs. 2(g-h)]. We found in the sample deposited at $T_D$ = 910 °C a root-mean-square (RMS) roughness $R_q$ equal to 0.55 Å, while, at higher $T_D$, graphene becomes definitely flatter, with the roughness being more than halved at 930 °C ($R_q$= 0.21 Å).

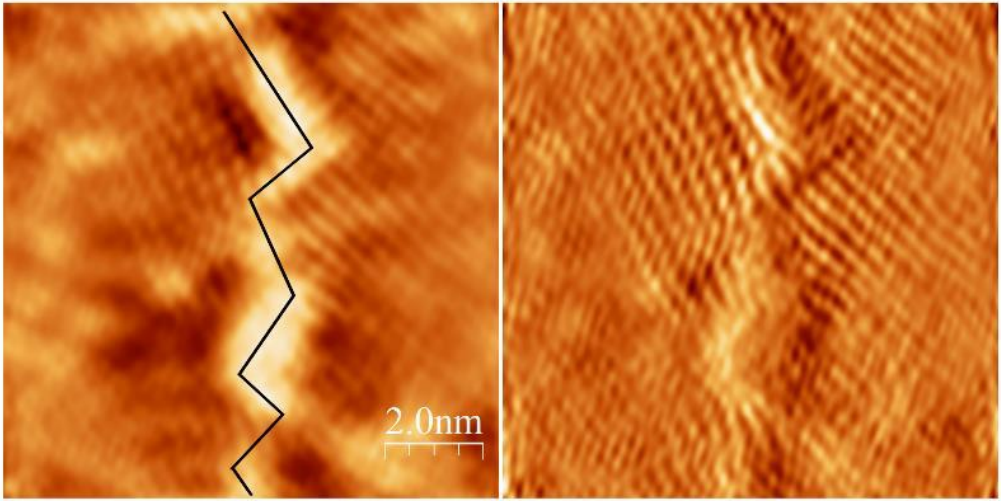

*Fig. 3. (left) STM image showing an example of grain boundary observed on the graphene sample deposited at $T_D= T_C= 910$ °C (U= -20 mV, I=1.0 nA). (right) Contrast-enhanced image obtained by evaluating the spatial derivative of the STM topography on the left.*

The STM investigation of the sample deposited at the lowest temperature ($T_D = 910$ °C) shows, in addition to the ripple morphology discussed above, the presence of several grain boundaries (GBs). An example is provided in Fig. 3 where, together with the STM topography, we also show a contrast-enhanced image obtained by evaluating its spatial derivative. The average height of GBs is ~0.2 nm, well matching previous reports [57]. GBs are known to be intervalley scattering centers, thus activating the Raman breathing mode of $sp^2$ C atoms forbidden in ideal graphene [58]. Their presence thus contributes to the D band intensity in the Raman spectrum of the low temperature sample [Fig.1(a)]. Notably, GB defects are absent in the graphene sample deposited at $T_D= 930$ °C. Further investigation of this latter sample is reported in Fig. 4. High-resolution STM images show a clear-cut honeycomb lattice extending over wide flat areas and resulting in an undistorted six-fold pattern obtained by fast-Fourier transform (FFT) [Figs. 4(a-c)]. On a larger scale [Fig. 4(d)], besides the presence of the Ge terraces and steps, STM images show small nuclei of graphene bilayer with a Moiré superlattice of ~4.38 nm periodicity [Fig. 4(e)]. From this value [59], we evaluate a misorientation angle between the two stacked graphene layers of about 3.2°. Considering that the bilayer nuclei are detected on a minority of STM sampling, we estimate the bilayer coverage to be < 2%. Interestingly, by changing the bias voltage, STM probes the Ge surface underneath the graphene.

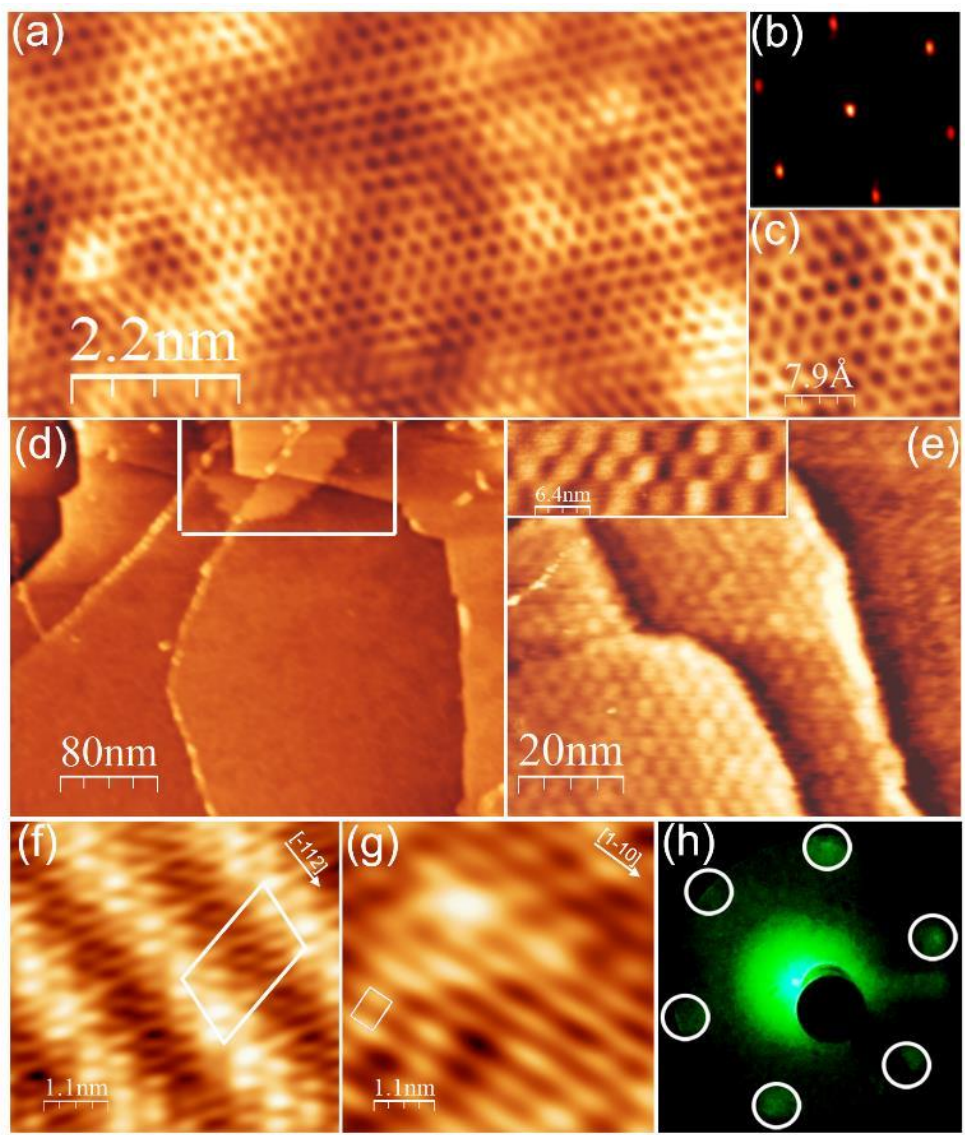

*Fig. 4. Graphene sample deposited at $T_D= T_C= 930$ °C: (a) honeycomb lattice of graphene monolayer. (b) FFT of the STM image and (c) enlarged view of graphene lattice. (d) large-scale STM picture of an area of the sample where a bilayer graphene is present in the center-top part of the image (marked). (e) enlarged views of the bilayer graphene showing a Moiré superlattice. On the right-most terrace, a monolayer graphene is present. STM images of (f) (6x2) reconstruction and (g) (1x1)-termination of the Ge(110) surface under the graphene monolayer. The surface unit cells are displayed. STM images*

*in panels (f) and (g) were acquired with U= -2.2 V, I= 0.5 nA, all the others with U= 50 mV, I=1.0 nA. (h) c-LEED image. Note that LEED was acquired without any UHV annealing after ex-situ transfer of the sample.*

In line with previous results [42], we find the coexistence of areas with different Ge surface patterns: zones showing a superstructure with a periodicity of about 1.7 nm [Fig. 4(f)] are alternated with areas where atomic rows with a much thinner spacing of ~0.55 nm are observed [Fig. 4(g)]. The former structure has been attributed to the (6x2) reconstruction of Ge(110) below the graphene. This reconstruction has not been observed for bare Ge(110), its being peculiar to the graphene/Ge system [42-44, 59]. Instead, the origin of structural motif in Fig. 4(g) is still being debated [42, 44]. The spacing observed between atomic rows is consistent with that of the unreconstructed (1x1) Ge(110) surface along the [001] direction (which is indeed the direction perpendicular to the rows). Therefore, it has been suggested that the surface pattern is due to the (1x1) bulk termination of Ge(110) stabilized by hydrogen present in the growth environment of CVD graphene [42]. However, it has been also proposed that the same structure can be interpreted as a diverse, novel surface reconstruction of Ge(110) strongly interacting with graphene and formed by post-growth UHV annealing to $T > 800$ °C [44]. We note that no such a treatment was performed on our samples; nonetheless, we mostly observe the typical pattern of Fig. 4(g), whereas the (6x2) reconstruction [Fig. 4(f)] is observed only locally, on less than 10% of the surface.

Structurally, the high-quality of the graphene film for $T_D$= 930 °C is confirmed by LEED. Figure 4(h) shows the conventional *c*-LEED pattern of the 930 °C sample. Only one set of hexagonally arranged diffraction spots are detected on the fluorescence screen, indicating that graphene is mostly single crystal. To obtain more quantitative information, we performed high-resolution LEED measurements by using a hemispherical electron analyzer (Fig. 5); the geometry of the setup is sketched in the inset. For $\phi= 0°$ the sample is aligned in the Γ-M direction (dark blue curve in Fig. 5). An ideal single-crystal graphene would show an individual reflection corresponding to $q_{//}= G_{\Gamma\text{-M}}$, which could also be visible in different diffraction orders. In our case this peak (with its second-order replica) is dominant, but we also observe an additional shallow feature. By changing the azimuth, we find that the intensity of this feature shows a maximum for $\phi= 30°$, i.e. when the azimuth is oriented along the Γ-K direction [light blue curve]. This reflection could be due to a minority of graphene domains with different crystalline orientation rotated by 30° with respect to the dominant one. The formation of such domains has been observed in the literature [24] as an additional subset of spots rotated by 30° in *c*-LEED. In contrast, in our case, these spots are not visible in *c*-LEED, indicating that this second orientational domain is clearly marginal. Indeed, from the intensity ratio between the Γ-M and Γ-K peaks in Fig. 5, we estimate the abundance of the minor orientation to be <10%. We remark that the Γ-K diffraction peak cannot be due to the second graphene layer, since the misorientation angle between the two stacked layers obtained from the Moiré superlattice is much lower than 30°.

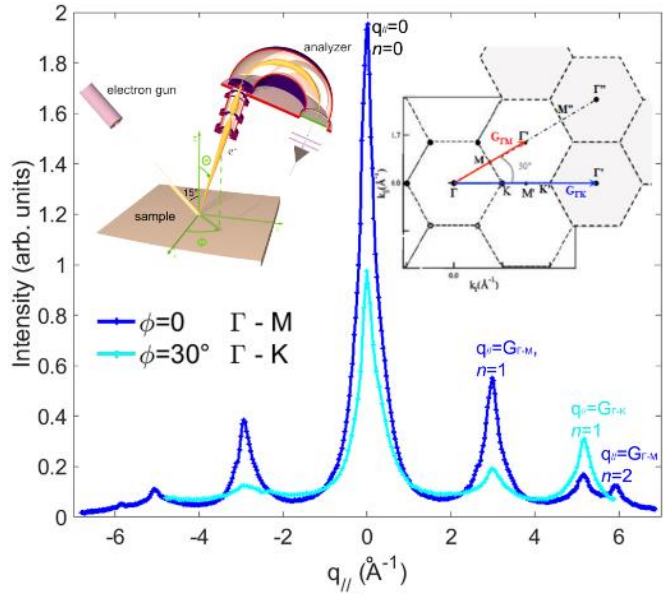

*Fig. 5. LEED spectra acquired at different azimuthal orientations on the graphene sample grown at at $T_D$ (= $T_C$)= 930 °C. The experimental geometry is sketched in the inset together with the reciprocal-space map of a graphene lattice. The electron energy used was 72 eV.*

Our data demonstrate that the quality of graphene/Ge(110) changes drastically when the deposition temperatures is slightly varied close to the melting point of Ge (938 °C). Namely, in our growth conditions, graphene of high quality is obtained at $T_D$= 930 °C but degrades significantly just 20 °C below this setpoint. In analogy with what has been recently observed on Ge(001) [29], we attribute this abrupt behavior to the incomplete surface melting of Ge where the negative Hamaker constant for long-range dispersion forces, typical of semiconductors, produces the stabilization of a quasi-liquid surface adlayer only in close proximity to the bulk melting point [48]. On Ge(110) the existence of this quasi-liquid adlayer only a few degrees below the Ge melting point has been experimentally proved [47]. Based on the temperature behavior observed, we believe that the sample deposited at 930 °C was grown at a temperature above the onset of the surface-melting regime, i.e. where a melted Ge adlayer is formed. Consistently, an abrupt change in the Ge topography occurs in this sample which develops large terrace widths due to the higher diffusivity and weakening of the in-plane bonding of surface Ge atoms in the melted regime. Graphene synthesis on a *quasi*-liquid Ge surface strongly differs from that occurring on a solid substrate. While growing, graphene tends to follow the potential energy minimum of the van der Waals interaction [60]: the presence of a mobile, liquid Ge layer hinders strong surface corrugations of the substrate, thus preventing the rippling of graphene. In addition, on a quasi-liquid substrate, the healing of defects is enhanced [29, 49, 61]. These mechanisms, acting above the onset of Ge surface melting, favor the formation of a flat, high-quality graphene film at high deposition temperature, as evidenced by Raman fingerprints (i.e. drop of D band intensity and increase of the $I_{2D}/I_G$ ratio). Conversely, the graphene growing at lower $T_D$ on a solid Ge surface needs to adapt to the uneven substrate corrugation: thus, it develops strong ripple-like corrugations and, eventually, a higher concentration of defects, like GBs. It was shown that the spontaneous formation of GBs is unlikely in case the graphene films is quasi free-standing and low interacting with the Ge substrate [19]. In our experimental conditions, the lowest degree of interaction between graphene and Ge (resulting in suppression of GBs) is clearly obtained at high deposition temperatures when the Ge surface is expected to be quasi liquid and provides a homogeneous support that allows the graphene layer to assemble and self-organize with a reduced interference from the support.

We believe that the onset of surface melting, and its relationship with graphene quality, also affects the graphene relaxation dynamics while cooling from the deposition to room temperature and hence the wrinkling. A requisite for wrinkles formation is that graphene can glide smoothly with little energetic expense on the substrate [56, 62]. This is indeed the case at high $T_D$ where the corrugation in the Ge substrate smoothens out due to the onset of surface melting, favoring thermal strain relaxation by wrinkling in this regime.

Finally, we suggest that the nucleation of embryos graphene bilayer observed at $T_D$= 930 °C is also in line with the formation of a quasi-liquid Ge surface layer which boosts the mobility of Ge atoms and their diffusional flow up to the surface of the first graphene layer, in line with what was observed on Ge(001) [22, 29].

Comparing our results with previous published works, we note that the synthesis of high-quality graphene with low D mode intensity was reported also at $T_D \cong$ 910 °C on both Ge(001) and Ge(110) [18-20, 30], so nominally below the onset of surface melting assuming the same temperature calibration (in our case, referred to a melted sacrificial sample). In these papers graphene was grown at much lower rates (at least, a factor of 10 [20, 30]) or with hydrogen absolute partial pressure being between 5 and >10 times higher than in our case [18-20]. Both these parameters are known to affect the quality of graphene. A slower growth rate may result in a smaller nucleation density and in an improved crystalline quality of the growing graphene islands [19]. Hydrogen was also found to play a primary role in determining graphene quality by etching away weak or defective carbon-carbon bonds [63]. Our results obtained using growth conditions characterized by faster growth rate combined with lower hydrogen absolute partial pressure highlight the role of the quasi-liquid Ge layer in the formation of high-quality graphene film. Interestingly, the abrupt increase of sublimation rate when entering the surface melting regime may, in our case, play a similar role as the hydrogen etching at higher absolute partial pressure, i.e. favoring the desorption of defective carbon species from the growing films. At the same time, the increased diffusivity of carbon species on the liquid Ge surface layer improves the crystalline quality of graphene [64].

As already pointed out, the Ge surface melting has a critical impact not only on graphene but also on the substrate morphology. Wide Ge terraces are observed below the graphene monolayer film deposited at $T_D$= 930 °C (Fig. 2). The low step density makes these Ge terraces potentially suitable as templates for graphene nanostructures such as nanoribbons. To this end, however, the graphene growth rate needs to be drastically reduced with respect to that we observed for $T_D$ in the 910-930 °C range. In order to decouple the effect of the Ge template morphology from the graphene growth process, we now explore the condition $T_C \neq T_D$. Therefore, after the pre-growth annealing of the Ge substrate to $T_C$= 930 °C we decreased the temperature and then performed the CVD deposition at $T_D$= 890 °C for a deposition time of 2 hours. Following this recipe, we obtain wide Ge terraces with widths of several hundreds of nanometers [Fig. 6(a)], closely resembling those obtained for $T_C=T_D$= 930 °C in Fig 2(f). The striking difference is that the Ge surface in Fig. 6 is mostly uncovered, appearing therefore mostly oxidized after transfer *ex situ* to the STM chamber [inset of Fig. 6(b)]. On the Ge surface, we only observe sparse ribbon nanostructures having widths of ~5 nm, typical length above 40 nm. By inspecting the large-scale STM image in Fig. 6(a), it is evident that the alignment of the ribbon features on Ge(110) is not stochastic. We find that the ribbon axis is aligned along specific high-symmetry directions of the Ge(110) surface, i.e. the [-112] and the [1-10] forming a relative angle of ~54° [42, 43]. Thanks to mirror symmetry of Ge(110) surface with respect to the [-110], the latter shows a twin orientation for ~74° azimuthal misorientation [41]. It is interesting to note that graphene islands with the armchair direction parallel to the [1-10] direction have been observed at the initial stage graphene growth on Ge(110) [19]. In addition, we note that the [-112] and the 74°-misaligned directions are also relevant for the graphene/Ge(110) interface, being related to the characteristic surface reconstruction of Ge(110) below graphene [42].

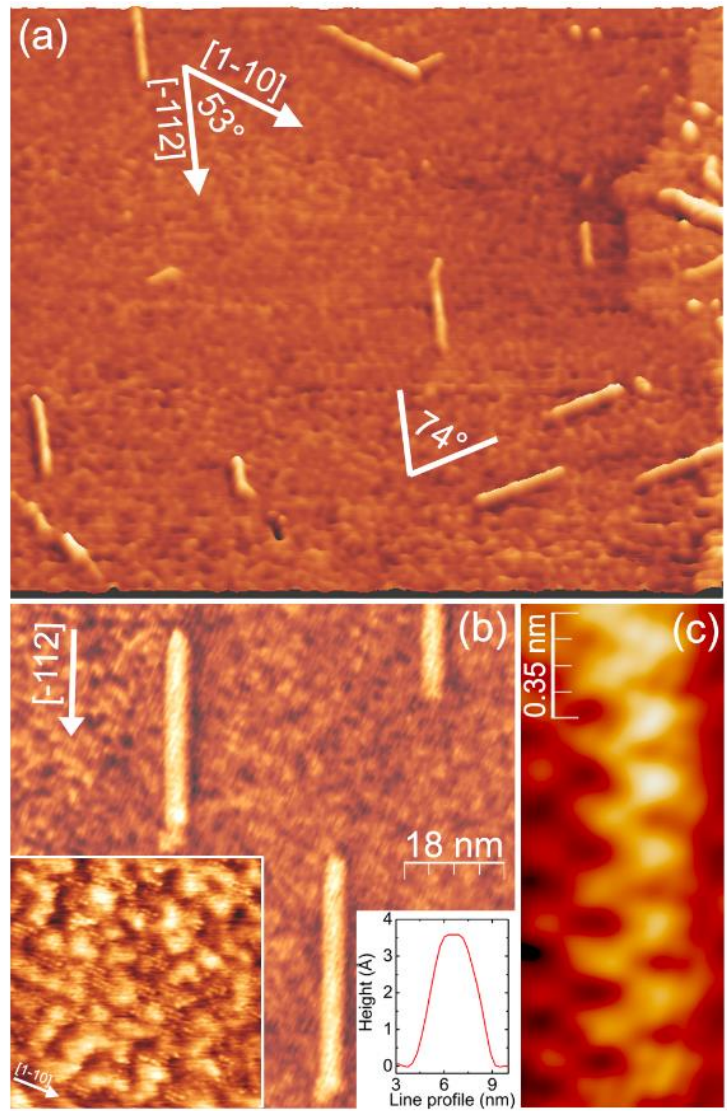

*__Fig. 6__. STM images of the sample annealed at $T_C$= 930 °C after deposition at $T_D$= 890 °C. (a, b) (U= -720 mV, I= 1.5 nA). The size of panel (a) is (325x 285) $nm^2$. In panel (b,) the left inset (25x 25 $nm^2$) shows an enlarged view of the oxidized Ge substrate, the right inset a STM line profile of a ribbon (c) STM pattern at the edge of ribbons (U= -300 mV, I= 1.7 nA).*

The height of the ribbon nanostructures is 3.3-3.6 Å [See inset Fig. 6(b)], a value being compatible with the out-of-plane spacing of graphene/Ge(110) [19]. Since the nominal hexagonal graphene lattice is not visible from atomic scale STM imaging which is dominated by complex scattering behaviors expected for such ultrathin widths [30, 31] [Fig. 6(c)], the unambiguous confirmation of the graphene composition of these structures would however require further optimization of the growth to increase their density and enable spectroscopic characterizations.

## 4. Conclusions

In conclusion, we have shown that both the deposition temperature of graphene and the annealing temperature during the cleaning step of Ge critically determine the results of CVD graphene growth on Ge(110). The quality of the continuous graphene film improves dramatically within a narrow range of deposition temperatures only very close to the melting point of Ge. We attribute this abrupt temperature dependence to the formation of a quasi-liquid Ge surface that favors the formation of high-quality graphene, thanks to the increased diffusivity and sublimation rate on the liquid Ge surface which promotes more effective diffusion of carbon species and desorption of the defective ones from the growing graphene films. This healing effect enabled by the Ge surface melting is particularly relevant for achieving high quality graphene at fast growth rates and lower hydrogen absolute partial pressure. In addition, we have found that the presence of the quasi-liquid Ge adlayer favors the formation of wide terraces and a low density of surface steps on the Ge(110) surface which appears to be promising as a template for the growth of ribbon nanostructures.

**Acknowledgements**

The Lime Laboratory of Roma Tre University is acknowledged for technical support.